\documentclass[10pt]{article}

\usepackage{amsmath}
\usepackage{amssymb}

\usepackage{graphicx}

\usepackage{cite}

\usepackage{color} 

\usepackage[labelfont=bf,labelsep=period,justification=raggedright]{caption}

\makeatletter
\renewcommand{\@biblabel}[1]{\quad#1.}
\makeatother

\date{}

\newcommand{\kb}{k_{\mathrm B}}

\usepackage{color}

\begin{document}

\begin{flushleft}
{\Large
\textbf{Thermodynamic costs of information processing in sensory adaptation}
}
\\
Pablo Sartori$^{1,\ast}$, 
L\'eo Granger$^{2}$, 
Chiu Fan Lee$^{3}$,
Jordan M. Horowitz$^{4}$
\\
\bf{1} Max Planck Institute for the Physics of
Complex Systems, Dresden, Germany
\\
\bf{2} Departamento de F\'isica At\'omica, Molecular y
Nuclear and GISC, Universidad Complutense de Madrid, Madrid, Spain
\\
\bf{3} Department of Bioengineering, Imperial College London, South Kensington Campus, United Kingdom
\\
\bf{4} Department of Physics, University of Massachusetts at Boston, Boston, MA, USA
\\
$\ast$ E-mail: pablosv@pks.mpg.de
\end{flushleft}

\section*{Abstract}
Biological sensory systems react to changes in their surroundings. They are characterized by fast response and slow adaptation to varying environmental cues. Insofar as sensory adaptive systems map environmental changes to changes of their internal degrees of freedom, they can be regarded as computational devices manipulating information. Landauer established that information is ultimately physical, and its manipulation subject to the entropic and energetic bounds of thermodynamics. Thus the fundamental costs of biological sensory adaptation can be elucidated by tracking how the information the system has about its environment is altered. These bounds are particularly relevant for small organisms, which unlike everyday computers operate at very low energies. In this paper, we establish a general framework for the thermodynamics of information processing in sensing. With it, we quantify how during sensory adaptation information about the past is erased, while information about the present is gathered. This process produces entropy larger than the amount of old information erased and has an energetic cost bounded by the amount of new information written to memory. We apply these principles to the \emph{E.~coli}'s chemotaxis pathway during binary ligand concentration changes. In this regime, we quantify the amount of information stored by each methyl group and show that receptors consume energy in the range of the information-theoretic minimum. Our work provides a basis for further inquiries into more complex phenomena, such as gradient sensing and frequency response.

\section*{Author Summary}

The ability to process information is a ubiquitous feature of living
organisms.
Indeed, in order to survive, every living being, from the
smallest bacterium to the biggest mammal,
has to gather and process information about its surrounding
environment.
In the same way as our everyday computers need power to function,
biological sensors need energy in order to gather and process this sensory information.
How much energy do living organisms have to spend in order to get
information about their environment?
In this paper, we show that the minimum energy required for a biological sensor to detect a change in some environmental signal is proportional to the amount of information processed during that event.
In order to know how far a real biological sensor operates from this minimum, we apply our predictions to chemo-sensing in the bacterium {\it Escherichia Coli} and find { that the theoretical minimum corresponds to a sizable portion of the energy spent by the bacterium.}

\section*{Introduction}

In order to perform a variety of tasks, living organisms
continually {respond} and adapt to their changing surroundings through
diverse electrical, chemical and mechanical signaling pathways, called
sensory systems~\cite{koshland1982amplification}.  In mammals,
prominent examples are the neurons involved in the visual, olfactory,
and  somatic systems~\cite{Gillespie1997,laughlin1989role,martelli2013intensity,abraira2013sensory}.
 But also unicellular organisms
lacking a neuronal system sense their environment: Yeast can sense
osmotic pressure \cite{Muzzey2009}, and {\it E. coli} can monitor
chemical gradients~\cite{shimizu2010modular},
temperatures~\cite{paster2008thermal} and pH~\cite{yang2012opposite}.
Despite the diversity in biochemical details, sensory adaptation
systems (SAS) exhibit a common behavior: long-term storage of the
state of the environment and rapid response to its changes
\cite{Smith2008}.  Intuitively, one expects that for these SAS to
function, an energy source -- such as ATP or SAM -- is required; {but is there a
fundamental minimum energy needed?}  To tackle this question, we
first relate a generic SAS to a {binary} information processing device,
which is tasked to perform fast information acquisition on the
environment ({response}) and to record subsequently the information into
its longer term memory (adaptation).  Since the foundational works of
Maxwell, Szilard and Landauer, the intimate relationship between
thermodynamic costs and information processing tasks has been
intensely
studied~\cite{Leff,Bennett:1982wx,Piechocinska2000,Dillenschneider2009,Sagawa2009,Granger2011,Granger2013}.
As a result, the natural mapping between a generic SAS and an
information processing device allows us to quantify the minimal
energetic costs of sensory adaptation.  

The idea of viewing biological processes as information processing
tasks is  not new~\cite{Bennett:1982wx,shimizu2010modular,Mehta2012}.  However, rationalizing sensory
adaptation is complicated by recent studies that have revealed that
 motifs in the underlying biochemical networks play a
fundamental role in the thermodynamic costs.  For instance, the steady
state of feedback adaptive systems must be dissipative, with more
dissipation leading to better adaptation~\cite{lan2012energy}, an
observation echoed in the analysis of a minimal model of adaptive
particle transport~\cite{Allahverdyan2013}.  Other studies have
suggested that some feedforward adaptive systems may require
dissipation to sustain their steady state~\cite{lan2013cost}, while
some may not~\cite{buijsman2012efficient,de2013unraveling}.
Furthermore, past studies \cite{Mehta2012,tostevin2009mutual} have
approached the notion of information by considering noisy inputs due
to stochastic binding, { a realm in which adaptation may not be relevant due to the separation of time-scales~\cite{sartori2011noise}}. 
Here, we develop a different approach that avoids these caveats by
considering a thermodynamically consistent notion of information that naturally incorporates the costs of sensing { in sensory adaptation.
Specifically, we derive a collection of universal bounds that relate the thermodynamic costs of sensing to the information processed.
These bounds reveal for the first time that for a generic SAS, measuring an environmental change is energetically costly [\eqref{eq:Imeas} below], while to erase the memory of the past is energetically free, but necessarily irreversible [\eqref{eq:Ierase} below].
By formalizing and linking the information
processing and thermodynamics of sensory systems, our work shows that
there is an intrinsic cost {of} sensing due to the necessity to process
information. 
}

To illustrate our generic approach, we study first a minimal
four-state feedforward  model and then a detailed ten-state feedback
model of {\it E. coli} chemotaxis.  Owing to the symmetry of its motif's topology
the four-state feedforward model
does not require energy to sustain its adapted state.
Instead, all the dissipation arises from information processing:
acquiring new information consumes energy, while erasing old
information produces entropy.  
By contrast, the \textit{E. coli} model
sustains its nonequilibrium steady state (NESS) by constantly
dissipating energy, a requirement for adaptation with a feedback topology~\cite{lan2012energy}.  
{ In this nonequilibrium setting, we generalize our thermodynamic bounds in order to pinpoint the additional energy for sensing over that required to maintain
the steady state.
We find with this formalism that in \emph{E.~coli} chemotaxis the theoretical minimum demanded by our bounds accounts for a sizable portion of the energy spent by the bacterium on its SAS.}

\section*{Results}

\subsection*{{ Universal} traits of sensory adaptation}

To respond and adapt to changes in an environmental signal $E$, a  SAS
requires a fast variable, the activity $A$; and a slow variable, the memory  $M$.
For example, in
\textit{E. coli} the activity is the conformational state of the receptor,
the memory the number of methyl groups attached to it, and the signal is the  ligand concentration~\cite{shimizu2010modular}.
Without loss of generality, we consider in the following all three variables normalized such that they only lie between $0$ and $1$, and
that the signal can only alternate between two values: a low value
 $0$ and a high value $1$.

As a result of thermal fluctuations,
the time-dependent activity $A_t$ and memory $M_t$ are stochastic variables.
Yet, the
defining characteristics of sensory adaptation are captured by their
ensemble averages $\langle A_t\rangle$ and $\langle
M_t\rangle$, both at the steady state and in response to
changes in the signal.

At a constant environmental signal $E=e$, the system relaxes to an
adapted $e$-dependent steady state,
which may be far from
equilibrium~\cite{lan2012energy}.
In this state, the memory is correlated with the signal, with an average value 
 close to the signal, $\langle M\rangle_{\rm
st}=|e-\epsilon_{\rm m}|$ where $\epsilon_{\rm m}$ is a small error. The
average
activity however is {\it adapted}, taking a value roughly independent
of the signal, 
$\langle A\rangle_{\rm st}=1/2\pm\epsilon_{\rm ad}$,
with adaption error $\epsilon_{\rm ad}$. 

Besides the ability to adapt, SAS are { also defined} by their { multiscale} response
to abrupt signal changes,
which is illustrated in Fig.~\ref{schemes}.
For example, given a sharp increase in the signal from $E=0$ to $1$
the average activity quickly grows from its adapted  value to a peak
$1-\epsilon_{\rm g}$ characterized by the gain error $\epsilon_{\rm
g}$. This occurs in a time $\tau_{\rm a}$, before the memory responds.
After a longer time $\tau_{\rm m}\gg\tau_{\rm a}$, the memory starts
to track the signal, and the activity gradually recovers to its
adapted value (see Fig. \ref{schemes}A).  
For a sharp decrease in the
signal, the behavior is analogous (see Fig. \ref{schemes}B).

We identify a SAS as any device that exhibits the described adapted
states for low and high signals ($0$ or $1$) and that reproduces the
desired behavior to abrupt increases and decreases in the signal (see Fig.~\ref{schemes}C for 
a cartoon biochemical example).
While SAS typically exhibit additional features (such as wide range
sensitivity~\cite{mello2007effects,Tu2008modeling}), they all { exhibit the universal features illustrated in} Fig.~\ref{schemes}.

\subsection*{{ Minimal SAS: equilibrium feedforward model}}

{ To facilitate the development of our formalism, we first present a minimal stochastic model of a SAS, where the activity $A$ and memory $M$ are binary variables ($0$ or $1$).
This model is minimal, since it has the least number of degrees of freedom (or states) possible and still exhibits the required response and adaptive behavior.
Treating the environmental signal $E$ as an external field that drives the SAS, the system can be viewed as evolving by jumping stochastically between its four states depicted in Fig.~\ref{fig:eqmodel}A.  The rates for activity $A$ transitions from $a^\prime \to a$ given $M=m$ at fixed $E=e$ are denoted $W_{a a^\prime}^{m}(e)$, and those for memory $M$ transitions from $m^\prime \to m$ given $A=a$ are $W^{m m^\prime}_a(e)$.

As an equilibrium model, it is completely characterized by a free energy function, which we have constructed in the Methods by requiring the equilibrium steady state to have the required signal correlations of a SAS,
\begin{equation}
	\label{eq:ener}
	F(a,m;e)=|e-m|(\Delta_{\rm m}+|e-a|\Delta_{\rm g}).
\end{equation}
 $\Delta_{\rm m} \approx k_{\rm B}T\ln \epsilon_{\rm m}^{-1}$ is the energy penalty for the memory to mistrack the signal, ensuring adaptation (with $T$ the temperature and $k_{\rm B}$ Boltzmann's constant). 
In fact, one can show that $\epsilon_{\rm ad}\approx \epsilon_{\rm m}/4$.
 $\Delta_{\rm g}\approx k_{\rm B}T \ln \epsilon_{\rm g}^{-1}$ is the penalty for the activity to mistrack the signal when $M\neq E$; it thus becomes relevant after a signal change, but before the memory adapts to the new signal, ensuring response.
In Figs.~\ref{fig:eqmodel}C and D the energy landscape $F(a,m;e)$ is represented for low and high signals (smaller radius corresponds to less probability and larger energy). 
Note that for fixed $E=e$, the adaptation error is zero when the energy penalty to misstrack the signal becomes large $\Delta_{\rm m} \to \infty$, the system's configuration is then $M=e$ and $A$ takes on the values $0$ and $1$ with equal probability.}
{ Finally, the dynamics are set by fixing the kinetic rates using detailed balance, \emph{e.g.},~$\ln W_{aa^\prime}^m(e)/W_{a^\prime a}^m(e)=-\left[F(a,m;e)-F(a^\prime,m;e)\right]/\kb T$, and then  choosing well-separated bare rates to set the timescale of jumps: $\omega$ for activity transitions and $k$ for memory transitions, with $\omega\gg k$, thereby enforcing the well-separated time-scales of adaptation.}

{When there is a change in the signal, this model exhibits response and adaptation
as characterized in Figs.~\ref{schemes}A and B}{~(verified in Figs.~1S and 2S)}, and relaxes towards a { \emph{dissipationless}} equilibrium steady state in which detailed balance is respected.
This is in contrast to previous studies on adaptive systems, which
demonstrated that maintaining the steady state for a generic feedback
system breaks detailed balance~\cite{lan2012energy,Allahverdyan2013}.
Our model, however, differs by its network
topology. As depicted in Fig.~\ref{fig:eqmodel}B, it is a mutually repressive feedforward (all rates
depend explicitly  on $E$, and the actions of $A$ and $M$ on each
other are symmetric). 
Similar topologies also underly recent suggestions for biochemical
networks that allow for adaptation with dissipationless
 steady states~\cite{buijsman2012efficient,de2013unraveling}.
 
\subsection*{Information processing in sensory adaptation}
Any sensory system that responds and adapts can naturally be viewed as an information processing device. In the steady state, information about the signal is stored in the memory, since knowledge of $M$ allows one to accurately infer the value of  $E$.
The activity $A$, on the other hand, possesses very little information about the signal, since it is adapted and almost independent of the signal.
When confronted by an abrupt signal change, the activity rapidly responds by gathering information about the new signal value.
As the activity decays back to its adapted value, information is stored in the memory.
However, to make room for this new information, the memory must decorrelate itself with the initial signal, thereby erasing the old information. 
Thus sensory adaptation involves measurement as well as erasure of information. 

{ To make this intuitive picture of information processing precise, let us focus on a  concrete experimental situation where the signal is manipulated by an outside observer.  
This is the setup common in experiments on {\it E. Coli} chemotaxis where the signal (the ligand
concentration) is varied in a prescribed, deterministic
way~\cite{segall1986temporal}.  
To be specific, the initial random signal $E_{\rm i}$ is fixed to an arbitrary value $e_{\rm i}$, either $0$ or $1$, with probability $p(e_{\rm i})$, and the system is prepared in the corresponding $e_{\rm i}$-dependent steady
state, characterized by the probability density $p_{\rm st}(a,m|e_{\rm i})$.
Then, at time $t=0$, the signal is randomly switched to $E_{\rm f}$ with final
value $e_{\rm f}=0,1$ (which may be the same as $e_{\rm i}$) according to the probability $p(e_{\rm f}|e_{\rm i})$.
The signal is held there while the system's time-dependent probability density $p_t(a,m|e_{\rm i},e_{\rm f})$, which conditionally depends on both the initial and final signals, irreversibly relaxes to the final steady state $p_{\rm st}(a,m|e_{\rm f})$.
During this relaxation correlations between the system and the final signal value $E_{\rm f}$ develop while the correlations with the past value $E_{\rm i}$ are lost.
As we will see, the measure of information that captures this evolution of correlations \emph{and}  naturally enters the thermodynamics of sensory adaptation is the mutual information between the system and the signal.

The \emph{mutual information} is an information-theoretic quantification of how much a random variable $U$ (such as the system) knows about another variable $V$ (such as the signal),
\begin{equation}\label{eq:mutualInfo}
I(U;V)=H(V)-H(V|U),
\end{equation}
measured} in nats \cite{Cover}. Here, $H(V)=-\sum p(v)\ln p(v)$ is the Shannon entropy, which is a measure of uncertainty. 
Thus, the mutual information measures the reduction in uncertainty of one variable given knowledge of the other.
Of note,  $I(U;V)\ge0$ with equality only when $U$ and $V$ are independent.

{ There are two key appearances of mutual information in sensory adaptation capturing how information about the present is acquired, while knowledge of the past is lost, which we now describe.
At the beginning of our experiment at $t=0$, the SAS is correlated with $E_{\rm i}$, simply because the SAS is in a $E_{\rm i}$-dependent steady state.
Thus there is an initial information $I(A_0,M_0;E_{\rm i})$ that the SAS has about the initial value of the signal $E_{\rm i}$.
The signal is then switched; yet immediately after, the SAS has no information about the new signal value $E_{\rm f}$, so $I(A_0,M_0;E_{\rm f})=0$.  Then for $t>0$ the SAS evolves, becoming correlated with $E_{\rm f}$, thereby gathering (or measuring) information $\Delta I_t^{\rm meas}=I(A_t,M_t;E_{\rm f})-I(A_0,M_0;E_{\rm f})\ge 0$}, which grows with time.
Concurrently it
decorrelates from $E_{\rm i}$, thus erasing information $\Delta I^{\rm
erase}_t=I(A_0,M_0;E_{\rm i})-I(A_t, M_t;E_{\rm i}|E_{\rm f})\ge 0$
about the old signal, which also grows with time.
This conditioning $I(A_t,M_t;E_{\rm i}|E_{\rm f})$ only takes into account direct correlations between $(A,M)$ and $E_{\rm i}$, excluding indirect ones through $E_{\rm f}$.

To illustrate this, we calculate
the flow of information  in the
non-disspative feedforward model for $p(e_{\rm i})=p(e_{\rm f}|e_{\rm i})=1/2$, which is a 1-bit operation (because $H(E_{\rm i})=\ln(2)\,{\rm nats}=1\,{\rm bit}$).  Figure~\ref{fig:infomeaser}A
displays the evolution of the measured information (in black),  which we
 decomposed as
\begin{equation}
\Delta I^{\rm meas}_t= I(M_t;E_{\rm f}) + I(A_t;E_{\rm f}|M_t)\equiv I^{(M)}_t+I^{(A|M)}_t,
\end{equation}
where $I^{(M)}$ (red) is the information stored in the memory and
$I^{(A|M)}$ (blue) in the activity.
We see the growth of $\Delta I^{\rm meas}$ proceeds first by a rapid ($t\sim\tau_{\rm a}$) increase as information is stored in the activity ($I^{(A|M)}$ grows) while the system responds, followed by a slower growth as adaptation sets in ($t\sim\tau_{\rm m}$), and the memory begins to track the signal.   
At the end, the system is adapted, and there is almost no information in the activity,
$I^{(A|M)}_{\infty}\approx0$. With the small errors we have, the information acquired reaches nearly the maximum value of $1$ bit, which is stored in the memory $\Delta I^{\rm meas}_\infty\approx I^{(M)}_\infty\approx1{\rm bits}$. Figure~\ref{fig:infomeaser}B shows the erasure of information, visible by the decrease of $I(A_t,M_t;E_{\rm i}|E_{\rm f})$ from an initial value of nearly one bit to zero when the system has decorrelated from the initial signal $E_{\rm i}$.


\subsection*{Thermodynamic costs to sensory adaptation}
{ We have seen that  through an irreversible relaxation, an SAS first acquires  and then erases information in the registry of the activity, followed by the memory.  
The irreversibility of these information operations is quantified by the entropy production, which we now analyze in order to pinpoint the thermodynamic costs of sensing.}  
Specifically, we demonstrate in Methods that for a system
performing sensory adaptation in response to an abrupt change in the
environment, the total entropy production can be partitioned in two
positive  parts: one caused by measurement ($\Delta S^{\rm meas}$) and
the other by  erasure ($\Delta S^{\rm eras}$). The second law thus becomes
\begin{equation}\label{eq:SmeasErase}
\Delta S^{\rm tot}_t=\Delta S^{\rm meas}_t+\Delta S^{\rm eras}_t\ge 0,
\end{equation}
{with the reference set to an initial state at $t<0$.} The erasure piece
\begin{equation}\label{eq:Ierase}
\Delta S^{\rm eras}_t=\kb\Delta I^{\rm eras}_t\ge 0,
\end{equation}
is purely {{ entropic in the sense that it contains no energetic terms.
It solely results} from the loss of information (or
correlation) about the initial signal.  By contrast, the energetics
are contained in the measurement portion,
\begin{equation}\label{eq:Imeas}
\Delta S^{\rm meas}_t=\kb\Delta H(A_t,M_t)-Q_t/T-\kb\Delta I^{\rm meas}_t\ge 0,
\end{equation}
where $\Delta H(A_t,M_t)=H(A_t,M_t)-H(A_0,M_0)$ is the change in Shannon entropy of the system
and { $Q_t=\int_0^t {\rm d}s\,\sum_{e_{\rm i},e_{\rm f}}p(e_{\rm i},e_{\rm f})\sum_{a,m}{\dot p_s}(a,m|e_{\rm i},e_{\rm f})F(a,m;e_{\rm f})$ is the average} heat flow into the system from the thermal reservoir.

A useful alternative formulation can be obtained once we identify the
internal energy $U_t$.  For example, in the equilibrium feedforward
model, a sensible choice is the average energy $U_t=\langle
F(A_t,M_t;E_t)\rangle$~\eqref{eq:ener}.  (Recall, that there is no unique
division into internal energy and work, though any choice once made is
thermodynamically consistent~\cite{Jarzynski2006b,Horowitz2007}.) By
substituting in the first law of thermodynamics $Q_t=\Delta U_t-W_t$, with $W_t$ the work,
we arrive at
\begin{equation}\label{eq:Imeas2}
W_t-\Delta {\mathcal F}_t\ge \kb T\Delta I^{\rm meas}_t.
\end{equation}
This equation shows how the measured information $\Delta I^{\rm meas}_t$
bounds the minimum energy required for sensing, which must be supplied
as either work $W_t$ or free energy ${\mathcal F}_t=U_t-\kb TH(A_t,M_t)$.
Thus, { \emph{to measure is energetically costly; whereas, erasure is
energetically free, but necessarily irreversible.}}  In particular, for
sensing to occur, the old information must be erased ($\Delta I^{\rm
erase}_t>0$), implying that the process is inherently irreversible,
\begin{align}
\label{eq:landauer}
\Delta S^{\rm tot}_t\ge \kb \Delta I^{\rm eras}_t>0.
\end{align}
Together \eqref{eq:Ierase} and \eqref{eq:Imeas2}  quantify the
thermodynamic cost of sensing an abrupt change in the environment by
an arbitrary sensory system.

We have demonstrated from fundamental
principles that sensing generically requires energy.  However,
\eqref{eq:Imeas2} does not dictate the source of that energy: It can
be supplied by the environment itself or by the SAS.  The distinction
originates because the definition of internal energy is not unique, a
point to which we come back in our analysis of 
\textit{E. coli} chemotaxis.

Using again our equilibrium feedforward model as an example, we apply our
formalism to investigate the costs of sensory adaptation.
Since this model sustains its steady state at no energy cost, the
ultimate limit lies in the sensing process itself.  We see this
immediately in Fig.~\ref{fig:eprdw} where we verify the inequalities
in \eqref{eq:SmeasErase} and \eqref{eq:Imeas2}.  Since $F$ in \eqref{eq:ener} is explicitly a function of the environmental signal $E$, the sudden change
in $E$ at $t=0$ does work on the system,  which is captured in
Fig.~\ref{fig:eprdw}A by the initial jump in $W$.  This work is
instantaneously converted into free energy $\Delta {\mathcal F}$ and is then consumed as the system responds and adapts in order to measure.  Thus, in
this example the work to sense is supplied by the signal (the
environment) itself and not the SAS, which is consistent with other
equilibrium models of SAS~\cite{de2013unraveling}.  Furthermore,
Fig.~\ref{fig:eprdw}B confirms that the erasure of information leads
to an irreversible process with net entropy production.
The bounds of \eqref{eq:SmeasErase} and \eqref{eq:Imeas2} are not tightly met in our model,
since we are
sensing a sudden change in the signal that necessitates a dissipative
response.  Nonetheless, the total entropy production and energetic
cost are on the order of the information erased and acquired.  This
indicates that these information theoretic bounds can be a limiting
factor for the operation of  adaptive systems. We now show that this is the case
for  {\it E. coli} chemotaxis, a  fundamentally different system as it operates far from equilibrium.

\subsection*{Extension to NESS and application to E. coli chemotaxis}
{ We have quantified the thermodynamic costs in any sensory adaptation system; however, for systems that break detailed balance and maintain their steady state far from equilibrium, \eqref{eq:Ierase} -- \eqref{eq:landauer} are uninformative, because of the constant entropy production. 
A case in point is {\it E. coli}'s SAS, which enables it to perform
chemotaxis by constantly consuming energy and producing entropy through the continuous hydrolysis of SAM.}

Nevertheless, there is a refinement of the second law for
genuine NESS in terms of the nonadiabatic $\Delta S^{\rm na}_t$ and
adiabatic $\Delta S^{\rm a}_t$ entropy productions, $\Delta S^{\rm
tot}_t=\Delta S^{\rm a}_t+\Delta S^{\rm na}_t$~\cite{Esposito2010}.
Crudely speaking, $\Delta {S}^{\rm a}$ is the entropy required to
sustain a nonequilibrium steady state and is never null for a genuine NESS; whereas $\Delta{S}^{\rm na}$ is the entropy produced by the transient time evolution.  
When the system satisfies detailed balance $\Delta{S}^{\rm a}_t=0$ always, be it at
its equilibrium steady state or not; when its surroundings change, the
entropy production is entirely captured by $\Delta{S}^{\rm na}_t$.
We can { refine our predictions for a NESS by recognizing that $\Delta{S}^{\rm na}_t$ captures the irreversibility due to  a transient} relaxation, just as $\Delta{S}^{\rm tot}_t$ does for systems satisfying detailed balance. {Analogously to Eqs. \eqref{eq:Imeas} and \eqref{eq:landauer}, we  derive} (see
Methods):
\begin{equation}\label{eq:nessImeas}
\kb\Delta H(A_t,M_t)-Q^{\rm ex}_t/T\ge \kb\Delta I^{\rm meas}_t,
\end{equation}
\begin{equation}\label{eq:nessIerase}
\Delta S^{\rm na}_t\ge\kb \Delta I^{\rm eras}_t\ge0.
\end{equation}
Here{, $Q^{\rm ex}_t=-\kb T\int_0^t{\rm d}s\, \sum_{e_{\rm i},e_{\rm f}}p(e_{\rm i},e_{\rm f})\sum_{a,m}{\dot p_s}(a,m|e_{\rm i},e_{\rm f})\ln p_{\rm st}(a,m|e_{\rm f})$} is the excess heat flow into the system,
roughly the extra heat flow during a driven, nonautonomous process over 
that required to maintain the steady state~\cite{Ge2010}.
{ As a result, it remains finite during an irreversible relaxation to a NESS, even though the NESS may  break detailed balance.}

\textit{E. coli} is a bacterium that can detect changes in the
concentration of nearby ligands in order to perform chemotaxis: the
act of swimming up a ligand attractor gradient. It is arguably the
best studied example of a SAS.   
At a constant ligand concentration $[L]$,
chemoreceptors in \textit{E. coli} -- such as the one in
Fig.~\ref{schemes}C -- have a fixed average activity, which through a
phosphorylation cascade translates into a fixed switching rate of the
bacterial flagellar motor.  When $[L]$ changes, the activity of the receptor $A$ (which is a binary variable labeling {two different receptor conformations}) increases on a
time-scale $\tau_{\rm a}\sim 1 {\rm ms}$.  On a longer time-scale
$\tau_{\rm m}\sim 10 {\rm s}$, the methylesterase CheR and
methyltransferase CheB  alter the methylation level of the receptor in
order to recover the adapted activity value.  In this way, the
methylation level $M$ (which ranges from none to four methyl groups {for a single receptor}) is a representation of the environment, acting as
the long-term memory (see diagram in Fig. \ref{chemoSchemes}A). One important difference with the previous equilibrium model is that the chemotaxis pathway operates via a feedback.  The memory is
not regulated by the receptor's signal, but rather by the receptor's
activity (see motif in Fig. \ref{chemoSchemes}B).  The implication is
that energy must constantly be dissipated to sustain the steady
state~\cite{lan2012energy}, thus \eqref{eq:nessImeas} and
\eqref{eq:nessIerase} are the appropriate tools for a thermodynamic
analysis.

There is a consensus kinetic model of \textit{E. coli} chemoreceptors
{\cite{keymer2006chemosensing,segel1986mechanism,shimizu2010modular,tu2013quantitative,Tu2008modeling}} whose
biochemical network is in Fig. \ref{chemoSchemes}A.  The  free energy landscape of the receptor coupled to its environment is
{\begin{align}\label{eq:F2}
F(a,m;[L])&={\Delta_{\rm m}}(a-\frac12)(m_0-m)+(a-\frac12)\ln\left[\frac{1+[L]/K_{\rm I}}{1+[L]/K_{\rm A}}\right]\\
&\equiv F_0(a,m)+V(a;[L])
\end{align}
with}  $\Delta_{\rm m}$ the receptor's characteristic energy, $m_0$ the reference methylation level, and $K_{\rm A/I}$ the active/inactive dissociation constants (values in Methods). In \eqref{eq:F2} the first term $F_0$ corresponds to the energy of the receptor, and the second $V$ comes from the interaction with the environment ({\it de facto} a ligand reservoir). The dynamics of this receptor consist of {thermal} transitions between the states with different activity, while transitions between the different methylation levels are powered by a chemical potential gradient $\Delta\mu=6k_{\rm B}T$ due to  hydrolisis of the methyl donor  SAM (see Methods). Continuous hydrolysis of SAM at the steady state sustains the feedback at the expense of energy, allowing accurate adaptation in the ligand concentration range $K_{\rm I}\ll[L]\ll K_{\rm A}$, see Fig.~\ref{chemoSchemes}B.

To begin our study, we  develop an equation analogous to
\eqref{eq:Imeas2}, which requires identifying the internal energy of
our system. 
As stated above, we consider the binding and unbinding of ligands as external stimuli, {and  thus define the internal energy as $U_t=\langle F_0(A_t,M_t)\rangle$. Using the excess heat $Q^{\rm ex}_t$, we consistently define the excess work through $W^{\rm ex}_t=\Delta U_t-Q^{\rm ex}_t$, analogous to the first law.} Upon substitution into \eqref{eq:nessImeas}
gives
\begin{equation}
W^{\rm ex}_t-\Delta {\mathcal F}_t \ge\kb T \Delta I^{\rm meas}_t,
\end{equation}
showing just as in
\eqref{eq:Imeas2} that measuring requires excess work and free energy.
Because here the
internal energy $U$ is \textit{not} a function of the ligand concentration,  $W^{\rm
ex}$ is not due to signal variation: It represents the energy expended by the 
cell to respond and adapt to the external chemical force.

In Fig.~\ref{chemoSchemes}C, we compare $W^{\rm ex}_t$ and $\Delta
{\mathcal F}_t$ to $\Delta I^{\rm meas}_t$ during a ligand change of $\Delta[L]\sim 10^2\mu{\rm M}$.  
The sudden change in $[L]$ produces a smooth, fast  ($\sim\tau_{\rm a}$) increase in the free energy as the activity transiently equilibrates with the new environment. 
The excess work driving this response comes mainly from the interaction with environment.
As adaptation sets in ($\sim\tau_{\rm m}$), the receptor utilizes that stored free energy, but in addition burns energy by the consumption of SAM.
Thus, in order to adapt the cell consumes the free energy stored from the environment, as well as additional excess work coming now mostly from the hydrolysis of SAM molecules. 
The inequality in \eqref{eq:Imeas2} with the measured information is satisfied at all times.

The energetic cost of responding and adapting to the ligand change is roughly $0.5k_{\rm B}T$, of which much has already been used by $t\sim\tau_{\rm m}=10s$. In comparison, the cost to sustain the chemotaxis pathway during this time is roughly $\sim 6k_{\rm B}T$ (see Methods). This means that the cost to sensing a step change is about 10\% of the cost to sustain the sensing apparatus at steady-state. During this process the cell measures (and erases) roughly $\sim0.3$ bits, less than the maximum of 1 bit despite its very high adaptation accuracy. 
This limitation comes from the finite number of discrete methylation levels, so that the probability distributions in $m$-space for large and low ligand concentrations have large overlaps{~(Fig.~S3).}
In other words, it is difficult to discriminate these distributions,  even though the averages are very distinct, which results in lower correlation between the methylation level and signal.
{ The minimal energetic cost associated to measuring these $\sim0.3$ bits ($\approx 0.2$ nats) is $0.2 \kb T$. {\it E. coli} dissipates roughly $0.5k_{\rm B}T$ during this process, thus the {energetic cost of sensory adaptation is slightly larger than twice its thermodynamic lower bound} ($2.5\approx 0.5/0.2$).}

We further explored the cost of sensing in \textit{E. coli} by examining
the net entropy production for ligand changes of different intensity. In Fig.~\ref{fig:chemosignals}A, we plot the amount of information erased/measured
for different step changes of the signal up to
$\Delta[L]\sim10^{5}\mu{\rm M}$ taking as lower base $[L]=50\mu{\rm M}$.  The green shading highlights the
region where adaptation is accurate ($\Delta [L]\ll K_{\rm A}$).  The  information erased is always below 1 bit and
saturates for high ligand concentrations, for which the system is not
sensitive.  The total entropic cost (that is, $\Delta S^{\rm
na}_\infty$) and its relation with the information erased appears in
Fig. \ref{fig:chemosignals}B. The dependence is monotonic{, and thus reveals a trade-off between information processing and dissipation in sensory adaptation. Notably,} for small acquisition of information (small ligand steps) it grows linearly with the information, an effect observed in ideal measurement systems~\cite{Granger2013}.


\section*{Discussion}

We have derived  {generic} information-theoretic bounds to sensory adaptation.
{ We have focused on response-adaptive sensory systems subject to an abrupt environmental switch.
This was merely a first step, but the procedure  we have outlined here only relies  on the validity of the second law of thermodynamics, and therefore can be extend to any small system affected by a random external perturbation to which we can apply stochastic thermodynamics, which is reviewed in~\cite{Seifert2012}.}

{ Our predictions are distinct from (although reminiscent of) Landauer's principle {\cite{Leff,Bennett:1982wx}},
which bounds the minimum energy required to reset an isolated  memory.  By contrast, the
information erased in our system is its correlations with the signal.
There is another important distinction from the setup of Landauer, and
more broadly the traditional setup in the thermodynamics of
computation~\cite{Leff} as well as the more recent advancements on the thermodynamics of information processing in the context of measurement and feedback~\cite{Sagawa2008,Sagawa2009,Sagawa2011b,Horowitz2010,Sagawa2013b,Horowitz2014,Cao2009,Barato2013,Ito2013}.  There the memory is reset by changing or
manipulating it by varying its energy landscape.  In our situation,
the erasure comes about because the signal is switched.  The loss of
correlations is stimulated by a change in the measured system -- that
is the environmental signal; erasure does not occur because the memory
itself is altered.
Also relevant is~\cite{Still2012}, which addresses the minimum dissipated work for a system to make predictions about the future fluctuations of the environmental signal, in contrast to the measured information about the current signal, which we have considered.}

Our results predict that energy is required to sense changes in the
environment, but do not dictate that source of energy. Our equilibrium feedforward model is able to
sense and adapt by consuming energy provided by the environment.
{\it E. coli}'s feedback, however, uses mostly external energy to respond,
but must consume  energy of its own to adapt. {The generic bounds here established apply to these two distinct basic topologies, irrespective of their fundamentally different energetics. For {\it E. coli}, }to  quantify to what extent $W^{\rm ex}$ is affected by SAM consumption and ligand binding, a more detailed chemical model is required in conduction with a partitioning of the excess work into distinct terms.

For a ligand change of $10^2\mu{\rm M}$, {in the region of high adaptation}, the information measured/erased is $\sim0.3$ bits. We observed that the corresponding average change in the methylation level for a chemoreceptor is $\sim0.75$, suggesting that a methylation level {can store} $\sim0.5$ bits for such 1-bit step response operations. {Despite the small adaptation error, information storage is limited by fluctuations arising from the finite number of discrete methylation levels. Receptors cooperativity, which is known to reduce fluctuations of the collective methylation level, may prevent this allowing them to store more information.} On the energetic side, we have shown that the cost of sensing these ligand changes per receptor is around 10\% of the cost of sustaining the corresponding adaptive machinery. We also showed that the energetic cost of binary operations is roughly twice beyond its minimum for large ligand changes, {in stark contrast with everyday computers for which the difference is orders of magnitude}. Taken together these numbers suggest that 5\% of the energy a cell uses in sensing is determined by  information-thermodynamic bounds, and is thus unavoidable.

Future work should include addressing sensory adaptation in more complex scenarios.
One which has recently aroused attention is 
 fluctuating environments, which so far has been addressed using trajectory information~\cite{Barato2013,diana2013mutual,Ito2013}.
However, under physiological conditions this is unlikely to play a significant role given the large separation of time-scales 
between binding, response, and adaptation \cite{sartori2011noise}. { Another scenario is a many bits step operation, in which instead of high and low signals a large discrete set of ligand concentrations is considered.} Frequency response and gradient 
sensing are also appealing \cite{Tu2008modeling}, since in them the system is in a dynamic steady state in which the memory is 
continuously erased and rewritten. Analysis of such scenarios is far from obvious, but the tools developed in this work constitute the first step in developing their theoretical framework.

\section*{Methods}

{\bf Kinetics of equilibrium { feedforward} model:} 
We determine a collection of rates that exhibit response and adaption
as in Fig.~\ref{schemes} by first decomposing the steady state
distribution as $p_{\rm st}(a,m|e)=p_{\rm m}(m|e)p_{\rm a}(a|m,e)$.
As a requirement to show adaptation, the memory must correlate
with the signal, which we impose by fixing $p_{\rm
m}(m|e)=\delta_{m,e} (1-\epsilon_{\rm m})+
(1-\delta_{m,e})\epsilon_{\rm m}$.  Next, in the steady state the
activity is $\langle A\rangle_{\rm st} \approx 1/2$, or since $A$ is
binary the probability $A=1$ is about $1/2$.  Recognizing that
$\epsilon_{\rm m}$ is small,  the average $\langle A\rangle_{\rm st}$
is dominated by adapted configurations with $M=e$.  Thus, adaption
will occur by demanding that $p_{\rm a}(1|0,0)=1/2-\epsilon_{\rm a}$
and $p_{\rm a}(1|1,1)=1/2+\epsilon_{\rm a}$, with a model parameter
$\epsilon_{\rm a}\ll 1$.  Finally, to fix the activity distribution
for non-adapted configurations, $M\neq e$, we exploit the time-scale
separation $\tau_{\rm a}\ll\tau_{\rm m}$.  In this limit, after an
abrupt change in the signal, the activity rapidly relaxes.  To
guarantee the proper response, we set $p_{\rm
a}(1|0,1)=1-\epsilon_{\rm g}$ and $p_{\rm a}(1|1,0)=\epsilon_{\rm
g}$.  Using the symmetry condition $p_{\rm st}(a,m|e)=p_{\rm
st}(1-a,1-m|1-e)$ we complete knowledge of $p_{\rm st}$.  The energy levels $F(a,m;e)$ are obtained using the equilibrium condition $F=-k_BT\ln p_{\rm st}$, where we choose as reference $F(0,0;0)=F(1,1;1)=0$. Equation~\eqref{eq:ener} is an approximation of this energy to lowest order in the small errors.  Finally, the
kinetic rates are obtained using either the approximate or exact
energy function, imposing detailed balance, and keeping two
bare rates, $\omega$ and $k$, for activity and memory
transitions{: $W^m_{aa^\prime}(e)=\omega e^{F(a^\prime,m;e)/\kb T}$  for activity transitions and $W_a^{mm^\prime}=ke^{F(a,m^\prime;e)/\kb T}$ for memory transitions.} \\

{ \noindent{\bf Information bounds on the thermodynamics of sensory adaptation:}}
The bounds in \eqref{eq:Ierase} and \eqref{eq:Imeas} follow from a
rearrangement of the second law of thermodynamics~\cite{Esposito2011}.
Consider a system with states $x$ [$(a,m)$ for SAS] { with signal-dependent (free) energy function $F(x;e)$} in contact with a thermal reservoir at temperature $T$.
The system is subjected to a random abrupt change in the signal. 
Specifically, the initial signal is a random
variable $E_{\rm i}$ with values $e_{\rm i}$ (which are $0,1$
in the main text), which we randomly change at $t=0$ to a new random
signal $E_{\rm f}$ with values $e_{\rm f}$.
{ For times $t>0$, we model the evolution of the system's stochastic time-dependent state $X_t$ as a continuous-time Markov chain.}

{ We begin our analysis by imagining for the moment that the signal trajectory is fixed to a particular sequence $(e_{\rm i},e_{\rm f})$. 
Then our thermodynamic process begins prior to $t=0$ by initializing the system in its $e_{\rm i}$-dependent steady state $p_{\rm st}(x|e_{\rm i})\propto e^{-F(x;e_{\rm i})/k_BT}$.
At $t=0$, the signal changes to $e_{\rm f}$ and remains fixed while the system's probability density $p_t(x|e_{\rm i},e_{\rm f})$, which conditionally depends on the \emph{entire} signal trajectory, evolves according to the master equation~\cite{VanKampen}
\begin{equation}\label{eq:master}
{\dot p}_t(x|e_{\rm i}, e_{\rm f})=\sum_{x^\prime\neq x} W_{xx^\prime}^{e_{\rm f}}p_t(x^\prime|e_{\rm i},e_{\rm f})-W_{x^\prime x}^{e_{\rm f}}p_t(x|e_{\rm i},e_{\rm f}),
\end{equation}
where $W_{xx^\prime}^{e_{\rm f}}$ is the signal-dependent transition rate for an $x^\prime \to x$ transition.
The transition rates are assumed to satisfy a local detailed balance condition, $\ln W_{xx^\prime}^{e_{\rm f}}/W_{x^\prime x}^{e_{\rm f}}=-(F(x;e_{\rm f})-F(x^\prime;e_{\rm f})/k_{\rm B} T$, which allows us to identify the energy exchanged as heat with the thermal reservoir in each jump.
Eventually, the system relaxes to the steady state $p_{\rm st}(x|e_{\rm f})\propto e^{-F(x;e_{\rm f})/k_B T}$ corresponding to the final signal value $e_{\rm f}$.

Since the signal trajectory is fixed, this process is equivalent to a deterministic drive by an external field, and therefore the total entropy production rate will satisfy the second law \cite{Esposito2011}
\begin{equation}\label{eq:2lawFixed}
	{\dot S}^{\rm tot}_t(e_{\rm i},e_{\rm f})=k_B\partial_t
	H (X_t|e_{\rm i},e_{\rm f})- {\dot Q}_t(e_{\rm i},e_{\rm f})/T\ge 0,
\end{equation}
where $\partial_tH(X_t|e_{\rm i},e_{\rm f})=-\sum_x {\dot p}_t(x|e_{\rm i},e_{\rm f})\ln p_t(x|e_{\rm i},e_{\rm f})$ is the rate of change of the Shannon entropy of the system conditioned on the entire signal trajectory; and 
\begin{equation}\label{eq:heat}
{\dot Q}_t(e_{\rm i},e_{\rm f})=\sum_x \dot{p}_t(x|e_{\rm_i},e_{\rm f})F(x;e_{\rm f})=-k_BT\sum_x{\dot p}_t(x|e_{\rm i},e_{\rm f})\ln p_{\rm st}(x|e_{\rm f})
\end{equation}
 is the
heat current into the system from the thermal reservoir given the
signal trajectory.  
Since \eqref{eq:2lawFixed} holds for any signal trajectory, it remains true after averaging over all signal trajectories sampled from the probability density $p(e_{\rm i},e_{\rm f})$:
\begin{equation}\label{eq:Stot}
{\dot S}^{\rm tot}_t=\kb \partial_tH(X_t|E_{\rm i},E_{\rm f}) - {\dot Q}_t/T\ge 0,
\end{equation}
with $H(X_t|E_{\rm i},E_{\rm f})=\sum_{e_{\rm i},e_{\rm f}}p(e_{\rm i},e_{\rm f}) H (X_t|e_{\rm i},e_{\rm f})$, and nonconditioned thermodynamic quantities, such as ${\dot Q}_t$, denotes signal averages.
 We next proceed by two judicious substitutions of the definition of the mutual information \eqref{eq:mutualInfo} that tweeze out the contributions from the measured and erased information.
First, we replace the Shannon entropy rate as $\partial_tH(X_t|E_{\rm i},E_{\rm f})=\partial_tH(X_t|E_{\rm f})-\partial_tI(X_t;E_{\rm i}|E_{\rm f})$, and then immediately repeat $\partial_tH(X_t|E_{\rm f})=\partial_tH(X_t)-\partial_tI(X_t;E_{\rm f})$.}
The result is a splitting of the total entropy production rate as $ {\dot S}^{\rm tot}_t={\dot S}^{\rm eras}_t+{\dot S}^{\rm meas}_t$, with one part due to erasure
\begin{equation}\label{eq:Serase}
	{\dot S}^{\rm eras}_t=-k_{\rm B}\partial_tI(X_t;E_{\rm i}|E_{\rm f})\ge0,
\end{equation}
and one due to measurement 
\begin{equation}\label{eq:Smeas}
	{\dot S}^{\rm meas}_t=k_{\rm B}\partial_t H(X_t)-{\dot Q}_t/T-k_{\rm B}\partial_t I(X_t;E_{\rm f})\ge 0.
\end{equation}  
The bounds in \eqref{eq:Ierase} and \eqref{eq:Imeas} follow by
integrating \eqref{eq:Serase} and \eqref{eq:Smeas}
from time $0$ to $t$.

To prove the positivity of \eqref{eq:Serase} and \eqref{eq:Smeas}, we use the definition of entropy and heat  to recast them in terms of a relative entropy
$D(f||g)=\sum_x f(x)\ln (f(x)/g(x))$~\cite{Cover} as {
\begin{align}
	\label{explicitSmesSer}
	{\dot S}^{\rm meas}_t&=-\kb\sum_{e_{\rm f}}p(e_{\rm f})\sum_{x}{\dot p}_t(x|e_{\rm f})\ln\frac{p_t(x|e_{\rm f})}{p_{\rm st}(x|e_{\rm f})}=-\kb  \sum_{e_{\rm f}}p(e_{\rm f}) \partial_t D[p_t(x|e_{\rm f})||p_{\rm st}(x|e_{\rm f})]\\
	{\dot S}^{\rm eras}_t&=-\kb\sum_{e_{\rm i},e_{\rm f}}p(e_{\rm i},e_{\rm f})\sum_{x}{\dot p}_t(x|e_{\rm i},e_{\rm f})\ln\frac{p_t(x|e_{\rm i},e_{\rm f})}{p_t(x|e_{\rm f})}=-\kb  \sum_{e_{\rm i},e_{\rm f}}p(e_{\rm i},e_{\rm f}) \partial_t D[p_t(x|e_{\rm i},e_{\rm f})||p_{\rm t}(x|e_{\rm f})].
\end{align}
Positivity then follows, since the relative entropy decreases whenever the probability density evolves according to a master equation, as in \eqref{eq:master}~\cite{Sagawa2013}.}

To arrive at \eqref{eq:nessImeas} and \eqref{eq:nessIerase}  for
genuine NESS, we repeat the analysis above applied to the average nonadiabatic
entropy production rate (cf.~\eqref{eq:Stot})
\begin{equation}
{\dot S}^{\rm na}_t=k_{\rm B}\partial_tH(X_t|E_{\rm i},E_{\rm f})- \frac{{\dot Q}^{\rm ex}_t}{T}\ge 0,
\end{equation}
 where ${\dot Q}^{\rm ex}_t=-k_{\rm B}T\sum_{e_{\rm i},e_{\rm f}}p(e_{\rm i},e_{\rm f})\sum_x \dot{p}_t(x|e_{\rm
i},e_{\rm f})\ln p_{\rm st}(x|e_{\rm f})$ is the excess heat flow into the system~\cite{Ge2010}, { taking special note that now $p_{\rm st}$ is the nonequilibrium stationary state and cannot be related to the energy, as in the equilibrium case above~\eqref{eq:heat}.}\\

\noindent{\bf Description of the chemotaxis model:}
The parameters for $F(a,m,s)$ in \eqref{eq:F2} are taken from
\cite{shimizu2010modular} for a Tar receptor: $K_{\rm I}=18.2\mu {\rm M}$,
$K_{\rm A}=3000\mu {\rm M}$, ${\Delta_{\rm m}}=2$, $m_0=1$.  The
kinetic rates are obtained using local detailed balance and
restricting to two characteristic time-scales. For $a$-transitions,
the rates are { $W_{aa^\prime}^m(e)=\tau^{-1}_{\rm a}\exp[ (a-a^\prime)(\Delta_{\rm
m}(m-m_{\rm o}) - e) /2]$}, with $\tau_{\rm a}=1{\rm ms}$ the typical
activation time. For $m$-transitions, the rates for active states are
{$W^{mm^\prime}_1=\tau^{-1}_{\rm m}(\delta_{m,m^\prime-1}+\delta_{m,m^\prime+1}\exp[-\Delta_{\rm m}/2+\Delta\mu])$, and for inactive states, $W^{mm^\prime}_0=\tau^{-1}_{\rm m}(\delta_{m,m^\prime+1}+\delta_{m,m^\prime-1}\exp[\Delta_{\rm m}/2-\Delta\mu])$.} Here,
$\Delta\mu=6k_BT$ is the chemical potential force for the hydrolyzation of a
SAM fuel molecule,  which occurs when a methyl group is added or
removed by CheR and CheB respectively~\cite{lan2012energy}, and at the steady state 
$\tau_{\rm m}\dot{S}^{\rm tot}_{\rm st}=\Delta\mu\approx6 k_{\rm B}T$.

\section*{Acknowledgments}

We are grateful to Y. Tu, D. Zwicker, R. Ma, R.G. Endres, G. Aquino, G. de Palo and S. Pigolotti for comments on this manuscript, and P. Mehta for helpful discussions.

\bibliography{sensbiblio,PhysicsTexts}

\newpage
\section*{Figure Legends}

\begin{figure}[!h]
\includegraphics[width=8.6cm]{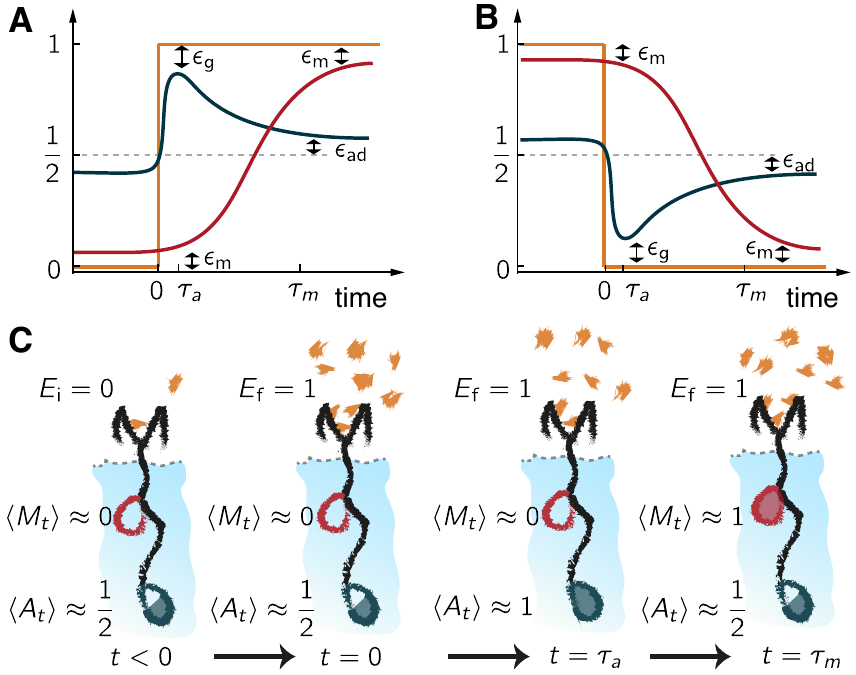}
	\caption{{\bf Generic traits of Sensory Adaptive Systems.} 
	(A/B) Typical time evolution of the average
	activity $\langle A_t\rangle$ (dark blue) and average memory
	$\langle M_t\rangle$ (red) of a SAS in response to an abrupt  increase
	or decrease in the signal $E$ (orange).  (C) Schematic
	states of a chemical receptor (black) embedded in a cell (light blue)
	during the four key phases of adaptation. {At $t<0$ the system is adapted; at $t=0$
	there is a sudden increase in the signal ligand concentration (orange flecks); at $t=\tau_\mathrm{ a}$ the receptor responds increasing its activity (full blue circle); and at time $t=\tau_\mathrm{ m}$ it is adapted (the memory is full, red; while the activity is half full blue).}
	\label{schemes}}
\end{figure}
\begin{figure}[!h]
\includegraphics[width=8.6cm]{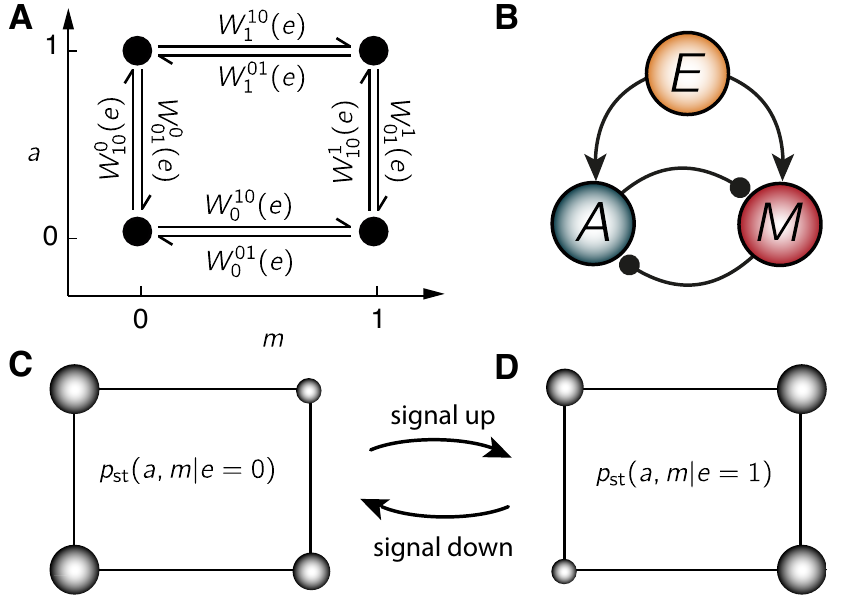}
	\caption{{\bf Equilibrium adaptation in a symmetric feedforward SAS.} (A) Reaction network of the four states in  activity, $a$, memory, $m$, space, with kinetic rates $W$ indicated for each transitions. (B) Topology of the model: feedforward with mutual inhibition. { For a fixed signal $E=e$, a sudden increase in the memory makes the average activity drop, and vice versa for activity changes. This symmetry of the topology, which is at the core of detailed balance, allows an equilibrium construction}. (C/D) Representation of steady state probabilities $p_\mathrm{ st}(a,m|e)$ for low/high $(0/1)$ signals using the $(a,m)$ space in (A). Wider state diameter represents higher probability, {thus lower energy}.
	\label{fig:eqmodel}}
\end{figure}
\begin{figure}[!h]
\includegraphics[width=8.6cm]{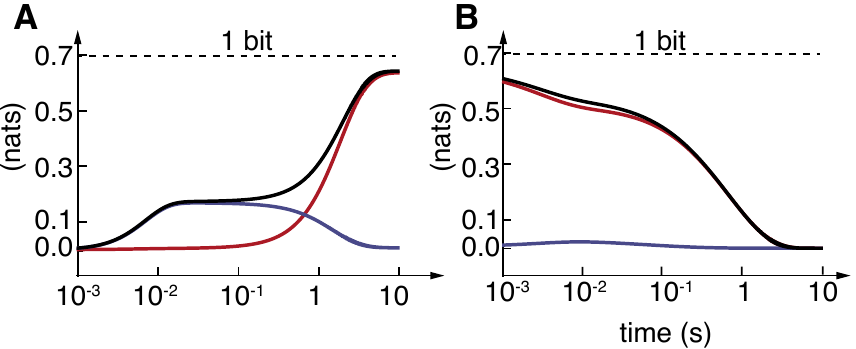}
\caption{{\bf Information measurement and erasure in sensory adaptation.} (A) Information acquired about the new signal as a function of
time. The information stored in the activity $I^{(A|M)}_t$ (dark blue) grows as the system responds, and then goes down as it adapts, when the information in the memory $I^{(M)}_t$ (red) grows. The total information measured  $\Delta I^\mathrm{ meas}_t$ { (black)} shows the effect of both.  (B)
Information lost about the old signal $I(A_t,M_t;E_\mathrm{ i}|E_\mathrm{
f})$ { (black)}, and its decomposition in memory { (red)} and activity { (blue)} information. Model parameters  are
$\epsilon_\mathrm{ x}=10^{-2}$ for x=a, m, g; $\omega=1/40 \,\mathrm{ s}^{-1}$ and $k=1/200\,\mathrm{ s}^{-1}$.
\label{fig:infomeaser}}
\end{figure}
\begin{figure}[!h]
\includegraphics[width=8.6cm]{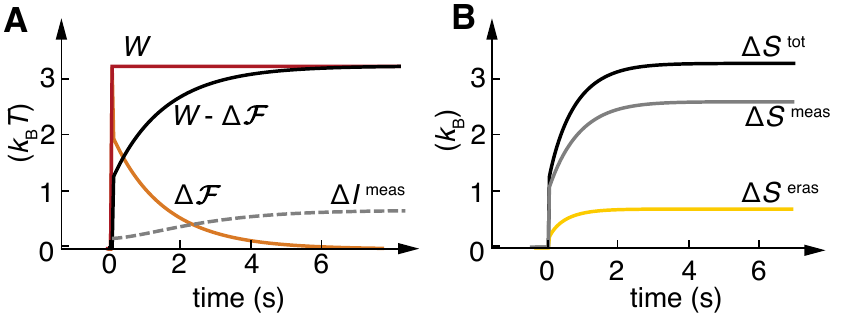}
\caption{{\bf Thermodynamics of adaptation in an equilibrium SAS.} (A) Energetic cost as a function of time given by the work $W$ provided by the environment (red), free energy change of the system $\Delta {\mathcal F}$ (orange), and 	dissipated work $W-\Delta {\mathcal F}$ (black), compared to the measured information $\Delta I^\mathrm{ meas}$ (grey dashed), which gives the lower bound at every time. (B) Total entropic cost $\Delta S^\mathrm{ tot}$ (black) and decomposition in measurement $\Delta S^\mathrm{ meas}$ (gray) and erasure $\Delta S^\mathrm{ eras}$ (yellow). Parameters  as in Fig.~\ref{fig:infomeaser}.
\label{fig:eprdw}}
\end{figure}
\begin{figure}[!h]
\includegraphics[width=8.6cm]{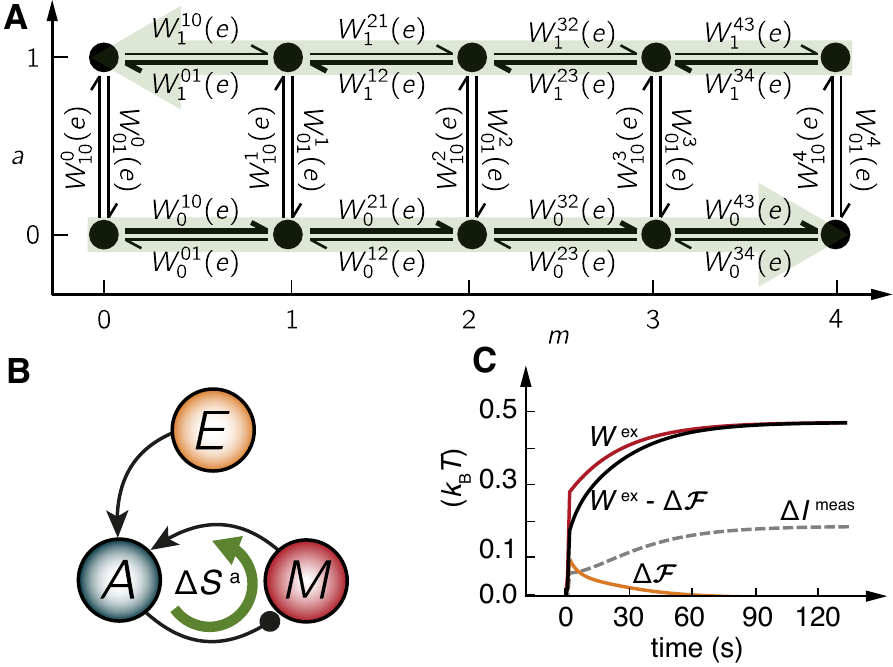}
\caption{{\bf Energetic costs of adaptation in an \emph{E. coli} chemotaxis SAS.} (A) Network representation of
the nonequilibrium receptor model with five methylation and two activity states. Green arrows represent the addition/removal of methyl groups driven by the chemical fuel SAM. (B) Corresponding negative feedback topology, displaying the dissipative energy cycle (green arrow) sustained by adiabatic entropy production, due to the consumption of chemical fuel. (C) Energetics of nonequilibrium
measurement in the chemotaxis pathway for a ligand concentration change of $\Delta L= 10^2\mu\mathrm{ M}$ (other parameters in Materials and Methods). The instantaneous change in ligand concentration performs chemical work on the cell, which increases its free energy $\Delta{\mathcal F}$ as the cell responds. To adapt, the bacterium has to provide excess work $W^{\rm ex}$ from its own chemical reservoir, the fuel SAM. \label{chemoSchemes} }
\end{figure}
\begin{figure}[!h]
\includegraphics[width=8.6cm]{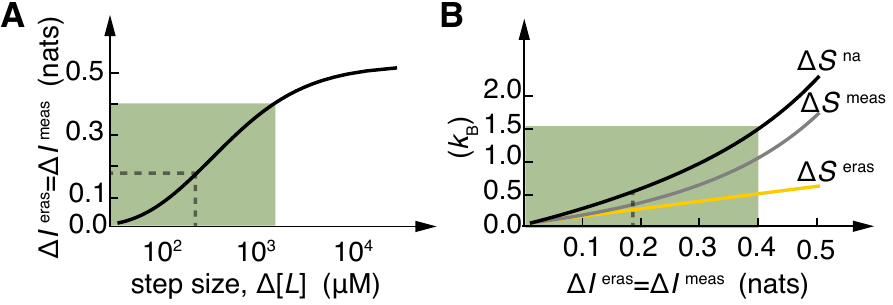}
\caption{{\bf Information-dissipation trade-off in \emph{E. Coli} chemotaxis.} (A) Relationship between information erased/acquired and size of the
signal increase. Shaded in green is the region of accurate adaptation
($\Delta [L]< K_\mathrm{ A}$). (B) Entropy
production as a function of information erased/acquired as step size is varied. The more information is processed by the cell the higher the entropic cost. Notice the linear scaling between dissipation and information for small information (small ligand changes).
\label{fig:chemosignals} Dashed lines refer to values  in Fig.~\ref{chemoSchemes}C. Parameters as in Methods.}
\end{figure}

\end{document}